\documentclass[conference, a4paper]{IEEEtran}
\usepackage{cite}
\usepackage{amsmath,amssymb,amsfonts}
\usepackage{algorithmic}
\usepackage{graphicx}
\usepackage{textcomp}
\usepackage{xcolor}
\usepackage{multirow}
\usepackage{listings}

\def\BibTeX{{\rm B\kern-.05em{\sc i\kern-.025em b}\kern-.08em
    T\kern-.1667em\lower.7ex\hbox{E}\kern-.125emX}}
\begin{document}

\title{Privacy-Preserving Data Quality Assessment for Time-Series IoT Sensors\\
% \thanks{}
}

% \author{
% \IEEEauthorblockN{Novoneel Chakraborty\IEEEauthorrefmark{1}\IEEEauthorrefmark{3}, Abhay Sharma\IEEEauthorrefmark{1}\IEEEauthorrefmark{4}, Jyotirmoy Dutta\IEEEauthorrefmark{1}\IEEEauthorrefmark{5}, Hari Dilip Kumar;\IEEEauthorrefmark{2}\IEEEauthorrefmark{6}}

% \IEEEauthorblockA{\IEEEauthorrefmark{1}Center of Data for Public Good, FSID, IISc}
% \IEEEauthorblockA{\IEEEauthorrefmark{2}Solvesustain}
% Bengaluru, Karnataka, India
% \\
% Email: \IEEEauthorrefmark{3}cnovoneel@gmail.com, \IEEEauthorrefmark{4}abhay.sharma@datakaveri.org, \IEEEauthorrefmark{5}jd222@snu.edu.in, \IEEEauthorrefmark{6}hari@solvesustain.com}

\author{\IEEEauthorblockN{Novoneel Chakraborty\IEEEauthorrefmark{1},
Abhay Sharma\IEEEauthorrefmark{1},
Jyotirmoy Dutta\IEEEauthorrefmark{1} and 
Hari Dilip Kumar\IEEEauthorrefmark{2}}
\IEEEauthorblockA{\IEEEauthorrefmark{1}Center of Data for Public Good, FSID, IISc}
\IEEEauthorblockA{\IEEEauthorrefmark{2}Solvesustain}
\IEEEauthorblockA{Email: cnovoneel@gmail.com, abhay.sharma@datakaveri.org, jd222@snu.edu.in, hari@solvesustain.com}
}

% \author{
% \IEEEauthorblockN{Novoneel Chakraborty}
% \IEEEauthorblockA{\textit{CDPG, FSID, IISc}\\
% Bengaluru, India \\
% cnovoneel@gmail.com}
% \and
% \IEEEauthorblockN{Abhay Sharma}
% \IEEEauthorblockA{\textit{Center of Data for Public Good} \\
% Bengaluru, India \\
% abhay.bits@gmail.com}
% \and
% \IEEEauthorblockN{Jyotirmoy Dutta}
% \IEEEauthorblockA{\textit{Center of Data for Public Good} \\
% Bengaluru, India \\
% jd222@snu.edu.in}
% \and
% \IEEEauthorblockN{Hari Dilip Kumar}
% \IEEEauthorblockA{\textit{Solvesustain} \\
% Bengaluru, India \\
% hari@solvesustain.com}
% }

\maketitle

\begin{abstract}
Data from Internet of Things (IoT) sensors has emerged as a key contributor to decision-making processes in various domains. However, the quality of the data is crucial to the effectiveness of applications built on it, and assessment of the data quality is heavily context-dependent. Further, preserving the privacy of the data during quality assessment is critical in domains where sensitive data is prevalent. This paper proposes a novel framework for automated, objective, and privacy-preserving data quality assessment of time-series data from IoT sensors deployed in smart cities. We leverage custom, autonomously computable metrics that parameterise the temporal performance and adherence to a declarative schema document to achieve objectivity. Additionally, we utilise a trusted execution environment to create a "data-blind" model that ensures individual privacy, eliminates assessee bias, and enhances adaptability across data types. This paper describes this data quality assessment methodology for IoT sensors, emphasising its relevance within the smart-city context while addressing the growing need for privacy in the face of extensive data collection practices. 
\end{abstract}


\section{Introduction}\label{intro}
The current era of computing can be characterised by the collection of vast amounts of data for applications such as decision making in governance, predictive analytics, healthcare, education and smart cities\cite{goudarzi},\cite{dewi},\cite{plageras}. Internet of Things (IoT) sensors are networked and interoperable devices that collect data about the environments in which they are deployed and play a critical role in the creation of data-driven applications\cite{kumar}. However, ensuring the quality of time-series data IoT sensors, especially in wide-scale deployments such as smart cities is difficult\cite{karkouch},\cite{cook},\cite{schultheis}. Manual assessment can be time-consuming and prone to bias, and traditional assessment methods may not adequately address privacy concerns with the increasing sensitivity of the data being collected, and may not be prudent to use in a scenario in which all parties operate with zero-trust\cite{freudiger}.

To address the need for a methodology that can adequately assess the quality of data generated by IoT sensors in an automated manner while preserving the privacy of individuals within that data, we propose a data quality assessment framework. We contextualise this framework for the smart-city paradigm, a complex use-case wherein large volumes of data are generated by IoT sensors with the potential for the data to contain Personally Identifiable Information (PII). We can thus subsequently distill our core objectives for the framework into three key aspects: \textbf{automatability}, \textbf{objectivity}, and \textbf{privacy}. 

To address automatability and objectivity, we propose custom, computable metrics that represent the most commonly cited data quality dimensions, making the assessment process automatable and non-reliant on comparison with other datasets. The performance of the sensor is measured in an objective manner with reference to its expected time-series behaviour, and the dataset is validated against a schema document that declaratively defines the structure and content of that dataset. 
To address the motive of privacy, we propose a model where we separate the data provider or the \textit{assessee}, from the assessment process, recognising that this can entail a conflict of interest. The assessment is carried out by a designated \textit{assessor} in a private, and 'data-blind' fashion, i.e. the assessor must not require any access to the data or the type of information the dataset contains in order to perform the assessment. To accomplish this, the assessment takes place within a trusted execution environment, in this case, a secure enclave. The benefits of this decentralised approach include preserving privacy, avoiding the introduction of assessee bias through the influence of preconceived notions, and increased flexibility and interoperability across data types, including Personally Identifiable Information (PII).

In this paper, we describe a novel method to privately assess the quality of time-series IoT data. The entirely objective nature of the methodology allows for end-to-end automation and scalable deployability of the system, while the 'data-blind' or private nature of the assessment is anticipatory of the increasing intricacies of sensitive data, their custodians, and the nature of their engagement with data consumers.

\section{Related Work}\label{relatedWork}

\subsection{Existing Approaches to Data Quality Assessment}
\subsubsection{Domain Specificity}
The problem of assessment of data quality has been tackled numerous times through the lens of multiple specific issues in various domains. The breadth of the existing methodologies is well documented by Cichy et Rass in their overview paper on data quality frameworks\cite{cichy}. This paper surveys twelve general-purpose data quality frameworks. In the survey, it is found that each of the frameworks considered different aspects or dimensions of data quality to be pertinent, depending on the use-case in which deployment of the methodology was envisaged. Wang describes four categories of information quality - intrinsic, accessibility, contextual, representational - and subsequently selects dimensions that are appropriate for each category, totalling to fifteen dimensions\cite{wang1}. Conversely, in Batini et al.’s work on heterogeneous data, two dimensions - accuracy and currency - are considered for their applicability to heterogeneous data specifically\cite{carlo}. 
It is apparent that appropriate dimension selection can make for a more tailored data quality assessment experience that focuses on the needs of a particular organisation or domain. This in turn, informs our work on selection of metrics that are bespoke to our specific use-case, i.e. private and automatable assessment of temporal IoT datasets.

\subsubsection{Types of Assessment}
Existing methodologies focus on either subjective, objective, or both forms of measurement, depending on parameters that they consider to be important. The AIM Quality (AIMQ) methodology proposed by Lee et al. for example, is an entirely subjective assessment, with a focus on an Information Quality Assessment questionnaire that serves to help an organisation perform a self-assessment\cite{lee}. Conversely, methodologies such as Hybrid Information Quality Management (HIQM) focus on a wholly objective nature to allow for a quantitative assessment, selecting the dimensions of accuracy, completeness, consistency, and timeliness\cite{cappiello}. The majority of methodologies fall somewhere along the spectrum bookended by these two extrema, combining an objective and subjective measurement to varying degrees. Task-Based Data Quality (TBDQ) is an example of such a methodology, where an initial, subjective assessment is performed through a questionnaire before being followed up by the computation of objective metrics\cite{vaziri}.

\subsection{Data Quality in the Context of Internet of Things}
Narrowing our focus to the domain of IoT sensors, the type of datasets become comparable to big data in terms of volume, velocity and variety,  making it difficult to have manual interventions in place in the form of subjective assessments. This is not a strictly followed rule, however. Fatouros et al.  propose a comprehensive architecture for DQA specifically for Industrial IoT (IIoT) applications wherein a subjective assessment is taken into consideration, albeit without any human intervention\cite{fatouros}. Similarly, Pipino presents an approach that combines both, subjective and objective assessments of data quality \cite{pipino}.

Methodologies such as the one proposed by Aquino et al. make an effort to evaluate the data values that a sensor is producing, by identifying outliers using the inter-quartile range (IQR) method\cite{aquino}. Similarly, Taylor and Loescher look at the values produced by a sensor to define a series of plausibility tests that assess the presence of outliers, and change in variance of values, amongst others\cite{taylor}. The automated assessment step is supplemented by human intervention.

For time-series data specifically, Gitzel proposes a hierarchical system of defining metrics, allowing for a drill-down from an elevated view of generally identified data quality issues. Gitzel also proposes a heatmap that showcases how the issues evolve with the associated timestamp allowing for a temporal view of the data quality\cite{gitzel}. 

In their survey paper on sensor data quality, Teh et al. summarise the problems that are generally faced in the domain of sensor data\cite{teh}. The most addressed are outliers, missing data, bias, drift, noise, constant value errors, uncertainty, and stuck-at-zero. There is a focus on the true-values of the data, the assessment of which lends a certain level of subjectivity. The determinant of quality can be dependent on the context and requirements of the data analysis.

\subsection{Defining Metrics}
In their paper on utility driven assessment of data quality, Even and Shankaranarayanan outline certain consistency principles for evaluation of data quality metrics\cite{even}. In the requirements for our methodology, we keep these consistency principles in mind, while leaving no scope for subjectivity in the form of human intervention, or availability of a reference dataset. 

\subsection{Private Assessment}
Freudiger et al. were the first to identify the class of problems in privacy-preserving data quality assessment and provide a series of protocols that rely on two-party secure computations for oblivious computation of metrics \cite{freudiger}. The privacy preserving protocols vary for the evaluation of each data quality metric, leading to a possibly complex deployment, but the methodology is not restricted to a specific domain or type of data. This differs from our approach, wherein we focus on an approach to private computation that is applicable regardless of the metric being computed, enabling the addition of more metrics to the methodology in the future, depending on the use-case. 

\subsection{Our Contribution}
The existing body of work consists of both subjective and objective assessment methods to assess DQ, and has been thoroughly contextualised for IoT devices. But there is a niche left unfilled when it comes to a universally private, automatable and objective methodology for the computation of data quality. To the best of our knowledge, this work addresses this specific gap. While we showcase our methodology using the smart-city scenario, our approach to the design of this methodology enables the creation of an end-to end solution that can be easily deployed on a large scale with metrics that can be customised to suit any domain or use-case.

\section{Private Assessment}\label{privateAssessment}
\subsection{The Case for Private Assessment}
The landscape of IoT applications is ever-expanding, and the smart city model is a compelling exemplar of one such deployment. Theoretical smart cities are a vast network of interconnected sensors and devices that are interwoven into the fabric of urban living. These sensors span domains from transportation and resource management to healthcare and public safety. These sensors may collect highly sensitive data about individuals and their activities, such as biometric data, civic data, social media data,  or health data\cite{vanzoonen}. This data can potentially be used by bad actors to compromise and intrude upon privacy rights. For individuals with privacy concerns, regulation and security mechanisms are important factors for the acceptance of IoT technologies\cite{psychoula}. 
The smart city paradigm lends itself to high stakes, where the imperative of utilising data to optimise urban function must be delicately balanced with the need to protect the privacy of those whose data is being collected. It acts as a microcosm of the implications and challenges of deployment of cyber-physical systems, making it an ideal case study to demonstrate the need and potential of a private data quality assessment technique.

\subsection{Stakeholders}
To avoid the potential for privacy breaches, provide enhanced security measures against malicious actors, and prevent the introduction of bias, it is prudent to adopt a decentralised, zero-trust framework between all the entities taking part in the data quality assessment. A zero-trust approach ensures a minimised attack surface, controlled data access, compliance with regulations, and secure collaboration between stakeholders, while decentralization maintains the integrity of the assessment, ensuring that the assessee's inherent bias cannot impact the outcome of the assessment. We consider an assessment model with three stakeholders:
\begin{itemize}
    \item Data Quality Assessee: The entity that owns the dataset that is to undergo a DQ assessment. The entity utilises a proxy resource server located on-premise, allowing the upload of sensitive data without exposure to other stakeholders.
    \item Data Quality Assessor: The entity that is providing the DQ assessment as a service. The DQ assessor uploads a specific configuration file matching the broad domain classification of the dataset (e.g., environment, urban mobility, waste management, energy).
    \item Secure Enclave: A dedicated, hardware-based, isolated system that provides a trusted environment for execution of the assessment computations. The enclave guarantees data confidentiality, and code and data integrity. Ownership, management, and maintenance of the enclave environment is assumed to be under the purview of the DQ assessor in this paradigm.
\end{itemize}

\subsection{Workflow}
The workflow consists of the following steps:
\begin{enumerate}
    \item The DQ Assessee uploads the dataset and schema to the proxy resource server, sharing the data domain with the assessor.
    \item The DQ Assessor uploads the required configuration file to the proxy resource server.
    \item The Secure Enclave loads the application code and receives an “address” to access the data and schema (encrypted by the proxy resource server) at the request of the DQ Assessor.
    \item The assessment is conducted within the trusted environment of the enclave.
    \item The generated output reports are encrypted and deposited back into the proxy resource server for access by the DQ Assessee.
\end{enumerate}

\begin{figure}[htbp]
    \centerline{\includegraphics[width=\columnwidth]{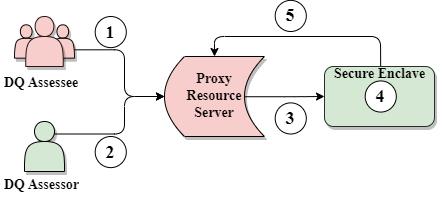}}
    \caption{Private Data Quality Assessment Workflow.}
    \label{DQPrivateAssessment}
\end{figure}

This workflow, shown in Fig. \ref{DQPrivateAssessment}, ensures the preservation of a zero-trust framework in which all users are granted resource access based on the principle of least privilege. This means that stakeholders are only given the access they need to perform authorised tasks and no more. Through this workflow, an entirely private assessment can be performed. 
It must be noted that since secure enclaves are isolated, hardware-based security protocols, they are generally implemented using dedicated hardware, which have some compute power and storage restrictions, implying a maximum permissible dataset size.

\section{Defining the Metric Requirements} \label{Requirements}
Using metrics to be able to rank datasets in order of their quality and usability will enable end users to make a better informed decision when selecting datasets for their needs. The metrics should also inform data providers about the performance of the sensors to some degree, so that remedial actions can be taken to mitigate potential data quality issues at the source. While not directly defining the bounds of privacy, the requirements should set the stage for private computation, ensuring that an end-to-end application can be developed to compute these metrics privately. The requirements thus derived are denoted by the set \textbf{R}, where the elements of set \textbf{R} are:

\begin{itemize}
    \item \textbf{R1 - Objective}: The metrics must be wholly objective, with no room for subjectivity. This is to eliminate the need for human intervention at any stage of the computation, and to ensure that the entire assessment pipeline is automatable.
    \item \textbf{R2 - Normalised}: The metrics must be normalised to a range between 0 and 1. This is done to ensure a consistent semantic interpretation of the scores.
    \item \textbf{R3 - Aggregatable}: The metrics must be aggregatable to ensure that a higher level view of data quality can be achieved after the assessment, i.e. they must have a common scale, unit of measurement, and support variable weightage to maintain some flexibility in terms of metric importance.
    \item \textbf{R4 - Dataset Agnostic}: The metrics should be applicable to all temporal (time-series) datasets.
    \item \textbf{R5 - Locally Computable}: The metrics should be computable with the available data and specified parameters only, and should not require any external information \textit{a priori} to complete the computation. This is to ensure that the assessment can take place in the black-box of a secure enclave and not require any additional information to compute.
\end{itemize}

\section{Defining the Data Quality Metrics}\label{DQMetrics}
\begin{figure}[htbp]
    \centerline{\includegraphics[width=0.5\textwidth]{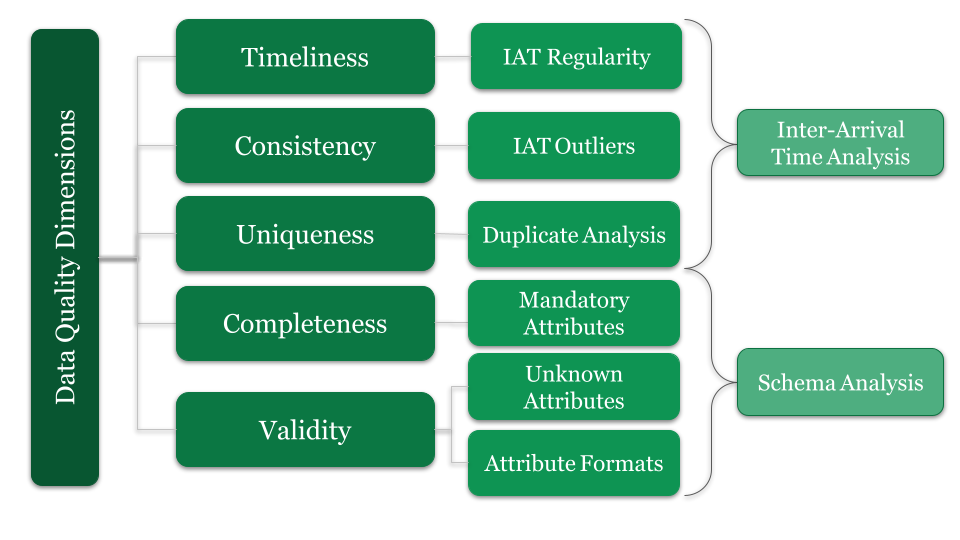}}
    \caption{Distillation of DQ Metrics from Dimensions.}
    \label{DQMetricsDistillation}
\end{figure}

The majority of literature around assessment of data quality makes references to the concept of dimensions. These dimensions are features or characteristics of the data that can be measured in order to assess the quality of the data\cite{askham}. There are a large number of these dimensions, as evidenced in \cite{cichy}, different frameworks recognise and co-opt different dimensions based on their focus and use-case. In our use-case, we wish to evaluate the quality of sensor data specifically and thus select those dimensions from the literature that are applicable. The most commonly referenced dimensions in the literature that are also applicable to sensor data quality are \textit{timeliness, consistency, uniqueness, validity, completeness, and accuracy}\cite{askham},\cite{wand},\cite{wang2}. The dimensions are distilled into metrics that meet the requirements list \textbf{R}. Fig. \ref{DQMetricsDistillation} shows the overall dimensions and the metrics that have been derived from them. 

Since requirement \textbf{R5} prohibits utilising external information to perform the computation, we cannot compute commonly cited metrics such as accuracy, resolution, and integrity. In each of these cases, additional information such as a reference dataset, or knowledge about nominal behaviour of the sensor and network are required.

The six derived metrics $\textbf{M} = \{M_{1},M_{2},M_{3},M_{4},M_{5},M_{6}\}$ can be grouped into two categories - assigned based on the specific aspect of the data they are assessing - Inter-Arrival Time [IAT] analysis and Schema analysis.

\subsection{Metrics for Inter-Arrival Time Analysis}
\label{metricsIAT}

First, we define three metrics $\{M_{1},M_{2},M_{3}\}$ that evaluate the performance of the sensor in terms of its inter-arrival time. IAT is broadly defined as the time between the arrivals of two consecutive events or entities.  In our case, the IAT is defined as the time elapsed between the arrivals of subsequent data packets from a sensor. 

\begin{figure}[ht]
    \centerline{\includegraphics[scale = 0.35]{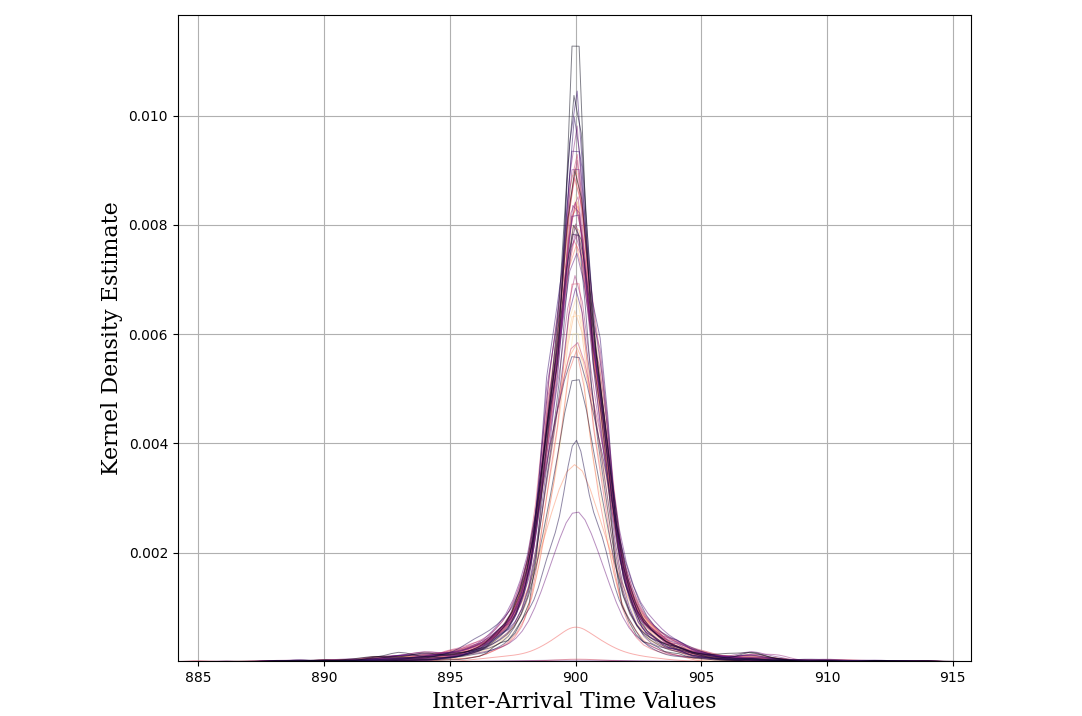}}
    \caption{Central tendency behaviour of IAT values for a group of sensors.}
    \label{NormalIAT}
\end{figure}

\begin{figure}[ht]
    \centerline{\includegraphics[width=\columnwidth]{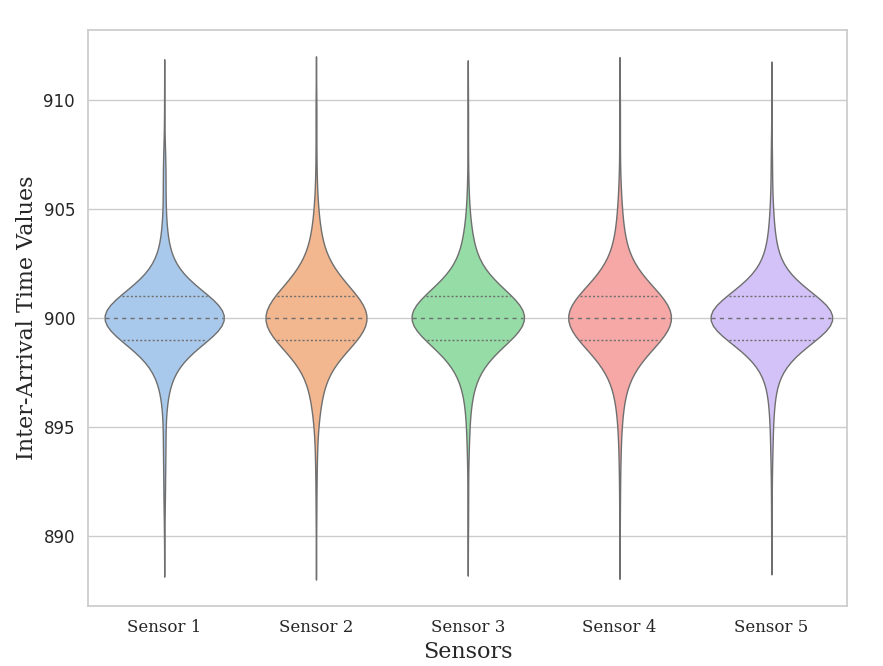}}
    \caption{Unimodal behaviour of sensor Inter-Arrival Time.}
    \label{UnimodalSensors}
\end{figure}

It is observed that the IAT values in data from smart city sensors are nominally distributed around a central tendency - the mode of the IAT values. Fig. \ref{NormalIAT} demonstrates this behaviour for data from a real-world air quality dataset, showing the clustering of the IAT values around a programmed reporting interval. This central tendency represents the expected or per-spec IAT for that sensor. This behaviour of the IAT values is consistent across types of IoT sensors, assuming that the sensors are of the same make and model, are programmed to report at the same interval, and are contributing data packets to the same dataset. There may be slight variations in the value of the central tendency depending on downtime, network congestion or latency. While bimodality is also observed, the majority of IATs cluster around a single central value indicating a predictable pattern of data packet generation and transmission. Fig. \ref{UnimodalSensors} illustrates the  unimodal distribution of IAT for sensors from real data of a city’s air quality monitoring system.

While these are our observations based on real-world temporal data from IoT sensors in numerous smart-city deployments, there may be other instances where the behaviour is different. For this paper, we evaluate the scenario in which the IAT values are unimodal.

\subsubsection{$M_{1}$: Inter-Arrival Time Regularity}
\paragraph{Definition}
IoT sensors can employ diverse triggers for data packet generation and transmission, including time-based, event-based, environmental, or user-initiated triggers, resulting in inter-arrival time distributions that may follow specific patterns. The regularity metric of the inter-arrival time is an evaluation of the uniformity of time intervals between receipt of consecutive data packets transmitted by IoT sensors. 

\paragraph{Computation}
In order to compute this metric we analyse the deviation of each inter-arrival time from the mode. To compute this deviation, we define:
\begin{equation}
RAE = \frac{\left |x_{i}-\bar{x}  \right |}{\bar{x}}
\end{equation}
Here, RAE is the Relative Absolute Error, \(x_{i}\) is the inter-arrival time, and x is the mode of the inter-arrival time. We consider an RAE value of 0.5 to be the crossover point between good and poor values of inter-arrival time, i.e. $RAE > 0.5$ is poor. We want to penalise the score proportionately to the RAE value, meaning the greater the RAE value, the greater the penalty. RAE is thus bound as RAE \(\in\) [0, \(\infty\)).

The metric score can thus be represented as \eqref{iatRegularity}.

\begin{equation}
\label{iatRegularity}
    M_{1}(x) = \frac{\sum\limits_{i:RAE_{i}\leq 0.5}^{}(1-2RAE_{i})}{\sum\limits_{i:RAE_{i}\leq 0.5}(1) + \sum\limits_{i:RAE_{i}>0.5}(2RAE_{i})}
\end{equation}

% \paragraph{Interpretation}
Equation \eqref{iatRegularity} represents the 'goodness' or ‘badness' of the inter-arrival time when compared to the modal value. The greater the difference between the IAT and the mode, the greater the penalty contribution to the regularity score for that IAT value. A value of 0.5 for RAE is chosen as the crossover point between ‘goodness’ and ‘badness’ of inter-arrival time. This follows from Lewis' interpretation of typical Mean Absolute Percentage Error (MAPE) values, where he describes the range of MAPE values $>50$ to be inaccurate forecasting\cite{lewis}. Since the computation of the MAPE includes the RAE, we can extrapolate this interpretation to the RAE as well. The value of 0.5 for RAE represents a window of values corresponding to:

\begin{equation}
    \bar{x} \pm \frac{\bar{x}}{2}
\end{equation}

This metric is particularly important for time-critical applications where a consistent and predictable inter-arrival pattern is desired.

\subsubsection{$M_{2}$: Inter-Arrival Time Outliers}
\paragraph{Definition}
The outlier metric of the inter-arrival time is an evaluation of the number of IAT values that show a significant deviation from the expected behaviour.

There are multiple ways to identify outliers in a dataset, and the choice of method is dependent on the independent characteristics of the dataset. In our case, we apply the modified Z-score method proposed by Iglewiscz and Hoaglin \cite{iglewicz}.

\paragraph{Computation}
Let the Median Absolute Deviation (MAD) of the data be defined as:

\begin{equation}
    m(x) = median_{i}(\left | x_{i} - \bar{x} \right |)
    \label{median}
\end{equation}

where $x_{i}$ is the observation for which the MAD is being computed and $\bar{x}$ is the mode of the data. We use the mode in place of the median as used in \cite{iglewicz} because we want to evaluate the deviation of the inter-arrival times from the mode, and we consider the mode to represent the expected behaviour of the dataset. Then the modified Z-score $Z_{i}$ is:

\begin{equation}
    Z_{i} = \frac{0.6745(x_{i} - \bar{x})}{m(x)}    
    \label{ModZScore}
\end{equation}

as follows from \cite{iglewicz}. In \cite{iglewicz}, it is recommended that observations with $\left | Z_{i} \right| > 3.5$ be classified as outliers, with variations to this cut-off value depending on the distribution of x. For our purposes, we will use this value as is to label inter-arrival time values as outliers. Once we have the number of outlier data packets, the metric can be computed as:

\begin{equation}
    M_{2}(x) = 1 - (\frac{\text{no. of outlier observations}}{\text{total no. of observations}})
\end{equation}

% \paragraph{Interpretation}
% The technique used captures how a specific observation’s behaviour deviates from the behaviour of the overall population.

\subsubsection{$M_{3}$: Duplicates}
\paragraph{Definition}
The duplicate analysis metric is a computation of the number of duplicate data packets in the dataset. These duplicates are identified by means of comparing a unique sensor identifier in each data packet with the time stamp of that data packet, under the assumption that no sensor can generate two or more data packets that have the same timestamp and attribute values.

\paragraph{Computation}
The duplicate analysis metric is computed using a simple ratio of the number of identified duplicate data packets to the total number of data packets.

\begin{equation}
    M_{3}(x) = 1 - (\frac{\text{no. of duplicate observations}}{\text{total no. of observations}})
\end{equation}

\subsection{Metrics for Schema Analysis}
Next, we define three metrics $\{M_{4},M_{5},M_{6}\}$ that are an analysis of a subset of the metadata that is provided along with the dataset. This is expected in the form of a schema, a document that delineates the structure of the dataset, such as the different types of attributes, the data types of each attribute (integer, float, string, etc.) as well as the ranges of the observations under each attribute. This document also provides the mandatory attributes that the dataset must contain, as well as a list of all the expected attributes in the dataset. This schema document is expected to be in JSON format. 

\subsubsection{$M_{4}$: Mandatory Attributes}
\paragraph{Definition}
The mandatory attributes metric checks whether the list of mandatory attributes defined in the data schema are all present in the dataset. This validation is performed for each data packet in the dataset.

\paragraph{Computation}
The following formula gives the metric score:

\begin{equation}
    M_{4}(x) = 1 - (\frac{\text{no. of observations with missing attributes}}{\text{total no. of observations}})
\end{equation}

% \paragraph{Interpretation}
% This metric is an indicator of the completeness of the dataset. If the schema defines a set of attributes that are mandatory or required to be present in each data packet, then the number of data packets with at least one attribute missing is computed. 

\subsubsection{$M_{5}$: Unknown Attributes}
\paragraph{Definition}
The unknown attributes metric is an evaluation of the number of data packets with attributes that are not specified in the schema in any capacity. 

\paragraph{Computation}
The metric is computed as a ratio of the number of data packets with at least one unknown attribute to the total number of attributes.

\begin{equation}
    M_{5}(x) = 1 - (\frac{\text{no. of observations with unknown attributes}}{\text{total no. of observations}})
\end{equation}

% \paragraph{Interpretation}
% This metric is computed by validating the data against the schema. The schema defines a set of attributes that are expected in the dataset. Any attributes that are outside this set of expected attributes are defined as unknown.

\subsubsection{$M_{6}$: Attribute Formats}
\paragraph{Definition}
The attribute format metric checks if the format of the data packets being evaluated matches the format defined in the data schema.

\paragraph{Computation}
The format adherence metric is computed by validating the data against the schema. The count of errors is incremented when the data type of an evaluated data packet does not match the data type specified in the data schema. The metric is computed as:

\begin{equation}
    M_{6}(x) = 1 - (\frac{\text{no. of observations with format errors}}{\text{total no. of observations}})
\end{equation}

\subsection{Summary of Metrics}
For all the metrics, 1 is a perfect score and 0 is the lowest possible score. The metrics thus defined are summarised in table \ref{MetricSummaryTable}.

% \usepackage{multirow}
% \usepackage{graphicx}
\begin{table}[htbp]
\caption{Summary of Data Quality Metrics}
\label{MetricSummaryTable}
\resizebox{\columnwidth}{!}{%
\begin{tabular}{|c|c|c|c|}
\hline
\textbf{Dimension} &
  \textbf{Metric} &
  \textbf{Description} &
  \textbf{\begin{tabular}[c]{@{}c@{}}Requirements \\ Met\end{tabular}} \\ \hline
Timeliness &
  $M_{1}$: IAT Regularity &
  \begin{tabular}[c]{@{}c@{}}Measures the uniformity of \\ time intervals between the \\ receipt of consecutive packets.\end{tabular} &
  All \\ \hline
Consistency &
  $M_{2}$: IAT Outliers &
  \begin{tabular}[c]{@{}c@{}}Evaluates whether there \\ are any outliers in \\ the inter-arrival times of \\ consecutive data packets.\end{tabular} &
  All \\ \hline
Uniqueness &
  $M_{3}$: Duplicate Analysis &
  \begin{tabular}[c]{@{}c@{}}Assesses the percentage \\ of duplicate data packets \\ present in the dataset.\end{tabular} &
  All \\ \hline
Completeness &
  $M_{4}$: Mandatory Attributes &
  \begin{tabular}[c]{@{}c@{}}Checks whether all the \\ required attributes defined in \\ the schema are present.\end{tabular} &
  All \\ \hline
\multirow{5}{*}{Validity} &
  $M_{5}$: Unknown Attributes &
  \begin{tabular}[c]{@{}c@{}}Checks for any additional \\ attributes in the dataset \\ beyond the required attributes \\ defined in the schema.\end{tabular} &
  All \\ \cline{2-4} 
 &
  $M_{6}$: Attribute Formats &
  \begin{tabular}[c]{@{}c@{}}Assesses the extent to which \\ the data adheres to the \\ expected format as defined \\ in the data schema.\end{tabular} &
  All \\ \hline
\end{tabular}%
}
\end{table}

The full list of requirements, \textit{{R1 - Objective, R2 - Normalised, R3 - Aggregatable, R4 - Dataset agnostic, R5 - Locally Computable}}  are thus met for each of the metrics. A detailed description of how the requirements are met are beyond the scope of this paper, but can be evaluated independently by the reader.

\section{Deployment}\label{deployment}
This DQ assessment methodology has been deployed and is currently in use by a smart-city data exchange that acts as an intermediary between a data provider, such as a civic body, and a data consumer, such as an application developer. The data exchange facilitates a consent-based exchange of data between these two entities through a platform based on open APIs and open data models. The data exchange operates and maintains the secure enclave environment for the purposes of private computation, and the DQ assessment takes place within this trusted execution environment. The DQ reports generated are shared with city administrators and government agencies to assist in diagnosis of DQ issues that may have a bearing on the health of the overall data ecosystem.

\section{Conclusions}\label{conclusions}
\subsection{Summary}
In this paper, we have presented a privacy-preserving data quality assessment for time-series IoT sensors. We created custom metrics that are representative of the most commonly cited dimensions in the literature of data quality assessment, ensuring that these metrics are objective and can be computed in an automated fashion. We then described a privacy-preserving architecture for deployment of this tool following a zero-trust model.

\subsection{Limitations}
This methodology is currently limited by certain design choices and hardware-based restrictions. It is designed for static datasets and cannot be currently applied to real-time streaming data. This restricts its applicability in scenarios where continuous monitoring and evaluation are critical. 
Additionally, the allocated hardware resources of the secure enclave is a primary determinant of the maximum permissible size of a static dataset.
Due to the nature of the computation within the secure enclave wherein only the data and the schema are used for computation, metrics such as accuracy, resolution, and integrity which rely on external references cannot be computed.
Finally, the effectiveness of metrics $M_{1}$ and $M_{2}$ hinge on the assumption that the IAT values exhibit unimodality and a clustering around the mode. Deviations from this behaviour could impact the accuracy of the assessment.

\subsection{Future Work}
In subsequent iterations of this tool and framework, we aim to build in functionality which will allow for a more real-time view of the data quality of a dataset or a sensor by assessing the most recent data from a dataset in a discrete time period. This will allow trends of data quality to develop over multiple such time periods, and can help assessees identify anomalous events and other behaviour that may not be captured in a static assessment. We will supplement this work with the development of a user-interface for the tool and a dashboard to analyse these trends. 
We also will further test and refine the metrics on more real-world datasets, to ensure broad compatibility and applicability across different domains. 

\section*{Acknowledgements}
% To be added for Camera-ready Submission
The methodology described in this paper was executed in three steps - ideation and research, development, and writing. These checkpoints would not have been possible without the contributions of certain members in the organisation. Bryan Paul Robert and Abhilash Venkatesh set up the secure enclave and iterated through the modifications required to the tool to ensure compatibility, and Dr. Anshoo Tandon reworked some of the metric computations and provided thorough feedback through the entire process.


\begin{thebibliography}{00}

\bibitem{goudarzi} S. Goudarzi, N. Kama, M. H. Anisi, S. Zeadally, and S. Mumtaz, “Data collection using unmanned aerial vehicles for Internet of Things platforms,” \textit{Computers and Electrical Engineering}, vol. 75, pp. 1–15, May 2019, doi: 10.1016/j.compeleceng.2019.01.028.

\bibitem{dewi} C. Dewi and R. C. Chen, “Decision making based on IoT data collection for precision agriculture,” in \textit{Studies in Computational Intelligence}, vol. 830, Springer Verlag, 2020, pp. 31–42. doi: 10.1007/978-3-030-14132-5\_3.

\bibitem{plageras} A. P. Plageras, K. E. Psannis, C. Stergiou, H. Wang, and B. B. Gupta, “Efficient IoT-based sensor BIG Data collection–processing and analysis in smart buildings,” \textit{Future Generation Computer Systems}, vol. 82, pp. 349–357, May 2018, doi: 10.1016/j.future.2017.09.082.
  
\bibitem{kumar} S. Kumar, P. Tiwari, and M. Zymbler, “Internet of Things is a revolutionary approach for future technology enhancement: a review,” \textit{Journal of Big Data}, vol. 6, no. 1, Dec. 2019, doi: 10.1186/s40537-019-0268-2.

\bibitem{karkouch} A. Karkouch, H. Mousannif, H. al Moatassime, and T. Noel, “Data quality in internet of things: A state-of-the-art survey,” \textit{Journal of Network and Computer Applications}, vol. 73. Academic Press, pp. 57–81, Sep. 01, 2016. doi: 10.1016/j.jnca.2016.08.002.

\bibitem{cook} A. A. Cook, G. Misirli, and Z. Fan, “Anomaly Detection for IoT Time-Series Data: A Survey,” \textit{IEEE Internet of Things Journal}, vol. 7, no. 7. Institute of Electrical and Electronics Engineers Inc., pp. 6481–6494, Jul. 01, 2020. doi: 10.1109/JIOT.2019.2958185.

\bibitem{schultheis} ]A. Schultheis et al., “Identifying Missing Sensor Values in IoT Time Series Data: A Weight-Based Extension of Similarity Measures for Smart Manufacturing,” in \textit{Lecture Notes in Computer Science (including subseries Lecture Notes in Artificial Intelligence and Lecture Notes in Bioinformatics)}, Springer Science and Business Media Deutschland GmbH, 2024, pp. 240–257. doi: 10.1007/978-3-031-63646-2\_16.
  
\bibitem{freudiger} J. Freudiger, S. Rane, A. E. Brito, and E. Uzun, “Privacy preserving data quality assessment for high-fidelity data sharing,” \textit{Proc. ACM Conf. Comput. Commun. Secur.}, vol. 2014-November, no. November, pp. 21–29, 2014, doi: 10.1145/2663876.2663885.

\bibitem{cichy} C. Cichy and S. Rass, “An overview of data quality frameworks,” \textit{IEEE Access}, vol. 7, pp. 24634–24648, 2019, doi: 10.1109/ACCESS.2019.2899751.

\bibitem{wang1} R. Y. Wang, “A Product Perspective on Total Data Quality Management,” \textit{Communications of the ACM}, vol. 41, no. 2, pp. 58-65, 1998. doi: 10.1145/269012.269022.

\bibitem{carlo} B. Carlo, B. Daniele, C. Federico, and G. Simone, “A Data Quality Methodology for Heterogeneous Data,” \textit{Int. J. Database Manag. Syst.}, vol. 3, no. 1, pp. 60–79, Feb. 2011, doi: 10.5121/ijdms.2011.3105.

\bibitem{lee} Y. W. Lee, D. M. Strong, B. K. Kahn, and R. Y. Wang, “AIMQ: A methodology for information quality assessment,” \textit{Information and Management}, vol. 40, no. 2, 2002. doi: 10.1016/S0378-7206(02)00043-5.

\bibitem{cappiello} C. Cappiello, P. Ficiaro, and B. Pernici, “HIQM: A methodology for information quality monitoring, measurement, and improvement,” \textit{Lect. Notes Comput. Sci. (including Subser. Lect. Notes Artif. Intell. Lect. Notes Bioinformatics)}, vol. 4231 LNCS, pp. 339–351, 2006, doi: 10.1007/11908883\_41.

\bibitem{vaziri} R. Vaziri, M. Mohsenzadeh, and J. Habibi, “TBDQ: A pragmatic task-based method to data quality assessment and improvement,” \textit{PLoS One}, vol. 11, no. 5, May 2016, doi: 10.1371/journal.pone.0154508.

\bibitem{fatouros} G. Fatouros, G. Makridis, A. Mavrogiorgou, J. Soldatos, M. Filippakis, and D. Kyriazis, “Comprehensive Architecture for Data Quality Assessment in Industrial IoT,” in \textit{Proceedings - 19th International Conference on Distributed Computing in Smart Systems and the Internet of Things}, DCOSS-IoT 2023, Institute of Electrical and Electronics Engineers Inc., 2023, pp. 512–517. doi: 10.1109/DCOSS-IoT58021.2023.00085.

\bibitem{pipino} L. L. Pipino, Y. W. Lee, and R. Y. Wang, “Data Quality Assessment,” \textit{Commun. ACM}, vol. 45, no. 4, pp. 211–218, Apr. 2002, doi: 10.1145/505248.506010.

\bibitem{aquino} G. R. Caldas De Aquino, C. M. De Farias, and L. Pirmez, “Hygieia: Data quality assessment for smart sensor network,” in \textit{Proceedings of the ACM Symposium on Applied Computing}, Association for Computing Machinery, 2019, pp. 889–891. doi: 10.1145/3297280.3297564.

\bibitem{taylor} J. R. Taylor and H. L. Loescher, “Automated quality control methods for sensor data: a novel observatory approach,” \textit{Biogeosciences}, vol. 10, no. 7, pp. 4957–4971, Jul. 2013, doi: 10.5194/bg-10-4957-2013.

\bibitem{gitzel} R. Gitzel, “Data quality in time series data: An experience report,” \textit{CEUR Workshop Proc}., vol. 1753, pp. 41–49, 2016.

\bibitem{teh} H. Y. Teh, A. W. Kempa-Liehr, and K. I. K. Wang, “Sensor data quality: a systematic review,” \textit{J. Big Data}, vol. 7, no. 1, pp. 1–49, Dec. 2020, doi: 10.1186/s40537-020-0285-1.

\bibitem{even} A. Even and G. Shankaranarayanan, “Utility-Driven Assessment of Data Quality,” \textit{Data Base Adv. Inf. Syst.}, vol. 38, no. 2, pp. 75–93, 2007, doi: 10.1145/1240616.1240623.

\bibitem{vanzoonen} L. van Zoonen, “Privacy concerns in smart cities,” \textit{Gov. Inf. Q.}, vol. 33, no. 3, pp. 472–480, Jul. 2016, doi: 10.1016/j.giq.2016.06.004.

\bibitem{psychoula} I. Psychoula, D. Singh, L. Chen, F. Chen, A. Holzinger, and H. Ning, “Users’ privacy concerns in IoT based applications,” \textit{Proc. - 2018 IEEE SmartWorld, Ubiquitous Intell. Comput. Adv. Trust. Comput. Scalable Comput. Commun. Cloud Big Data Comput. Internet People Smart City Innov. SmartWorld/UIC/ATC/ScalCom/CBDCom/IoP/SCI 2018}, pp. 1887–1894, 2018, doi: 10.1109/SmartWorld.2018.00317.

\bibitem{askham} N. Askham et al., “The Six Primary Dimensions for Data Quality Assessment,” Group, DAMA UK Working. Accessed: Feb. 23, 2022. [Online]. Available: https://www.dqglobal.com/wp-content/uploads/2013/11/DAMA-UK-DQ-Dimensions-White-Paper-R37.pdf

\bibitem{wand} Y. Wand and R. Y. Wang, “Anchoring data quality dimensions in ontological foundations,” \textit{Commun. ACM}, vol. 39, no. 11, pp. 86–95, Nov. 1996, doi: 10.1145/240455.240479.

\bibitem{wang2} R. Y. Wang and D. M. Strong, “Beyond Accuracy: What Data Quality Means to Data Consumers,” \textit{J. Manag. Inf. Syst.}, vol. 12, no. 4, pp. 5–33, 1996.

\bibitem{lewis} C. D. Lewis, \textit{Industrial and business forecasting methods: a practical guide to exponential smoothing and curve fitting}. London; Boston: Butterworth Scientific, 1982.

\bibitem{iglewicz} B. Iglewicz and Hoaglin David C., \textit{How to Detect and Handle Outliers}. The ASQC Basic References in Quality Control: Statistical Techniques, vol. 16, United States of America: American Society for Quality Control Statistics Division, 1993.
\end{thebibliography}
\end{document}